\documentclass[a4paper,11pt]{article}
\usepackage{pos}
\usepackage{cleveref}
\usepackage{wrapfig,lipsum}
\usepackage{hyperref}

\title{Neutron stars and Constraints for the Equation of State of Dense Matter}

\author[a]{Rajesh Kumar}
\author*[a]{Veronica Dexheimer}
\author[b]{Johannes Jahan}

\affiliation[a]{Department of Physics, Kent State University \\
800 E Summit St, Kent, OH, USA 44240}

\affiliation[b]{Department of Physics, University of Houston,\\
3507 Cullen Blvd Room 617, Houston, TX, USA 77204}

\emailAdd{vdexheim@kent.edu}

\abstract{
Neutron stars provide a natural laboratory for studying the properties of dense nuclear matter under extreme conditions. In this proceeding, we review our current understanding of dense isospin symmetric and asymmetric matter and neutron star physics. We focus on modern theoretical, experimental, and observational constraints, including first-principle calculations from lattice and perturbative Quantum Chromodynamics (QCD), as well as chiral effective field theory approaches at nuclear densities. From the experimental perspective, constraints on the equation of state arise from heavy-ion collisions, low-energy nuclear physics, and astrophysical observations, including neutron star masses, radii, and gravitational wave signatures from mergers. These multidisciplinary comparisons are crucial for bridging the gap between nuclear physics and astrophysical observations, in order to expand our knowledge of matter at supra-nuclear densities.}

\FullConference{
}

\begin{document}
\maketitle

\section{Introduction}

Neutron stars are among the most extreme astrophysical objects, with central baryon densities, $n_B$, reaching several times the nuclear saturation density, $n_{\rm sat}$. These compact remnants of massive stars provide an excellent environment to explore dense nuclear matter, where nuclei dissolve into hadrons (baryons, containing 3 valence quarks each), which may coexist with or transition into deconfined quark matter \cite{Baym:2017whm}. Understanding the equation of state (EoS) of neutron star matter is a fundamental challenge in nuclear and astrophysics, requiring insights from both theoretical and observational constraints.
At densities around $n_{\rm sat}$, strongly interacting matter is mainly composed of nucleons (protons and neutrons) bound by nuclear forces. In neutron stars, the number of neutrons is much larger than the one of protons, balanced by electrons to fulfill charge neutrality. This difference creates an isospin imbalance or asymmetry, which extends to matter containing heavier baryons (hyperons) or deconfined quarks. Various theoretical approaches have been developed to study the low temperature, effectively zero on the MeV scale, EoS across different $n_B$s. In the low-density regime, Chiral Effective Field Theory ($\chi$EFT) can be employed to calculate the EoS relevant around $n_B \sim n_{\rm sat}$ of neutron stars~\cite{Drischler:2013iza}. \textit{Ab initio} $\chi$EFT calculations allow for the study of the EoS at arbitrary isospin asymmetry within many-body perturbation theory or the Brueckner–Hartree–Fock approach~\cite{Logoteta:2016hxh}. Benchmark calculations with the first and second generation of $\chi$EFT Norfolk 2 and 3-body interactions have been performed to assess the uncertainties associated with studying the many-body Schrödinger equation~\cite{Piarulli:2019pfq}. These methods provide reliable constraints; however, their validity becomes uncertain at higher densities due to missing higher-order many-body interactions and non-perturbative QCD effects.

In the high-temperature ($T > 125$ MeV) and low baryon chemical potential $\mu_B$ regime, where the number of baryons (or quarks) is or almost is matched by the number of antibaryons (or antiquarks),  Lattice QCD (LQCD) can be applied. LQCD is a non-perturbative technique for solving the quantum chromodynamics (QCD) theory of quarks and gluons formulated on a grid of points in space and time, and it is a powerful first-principles approach for studying the thermodynamic properties of strongly interacting matter. LQCD calculations successfully describe the EoS at zero and imaginary baryon chemical potential $\mu_B$~\cite{Ratti:2018ksb}, but due to the infamous sign problem \cite{Troyer:2004ge}, direct simulations at real finite $\mu_B$ or $n_B$ are currently not feasible. Interestingly, this problem is not present for isospin chemical potential $\mu_Q$. Alternative methods, such as Taylor expansions in $\mu_B/T$ and analytical continuation from imaginary chemical potential, have been extremely successful in extending our knowledge of QCD to finite $\mu_B$~\cite{Borsanyi:2025dyp}. Nevertheless, these techniques are still limited to moderate $\mu_B$ and high $T$. 
In the extremely high $n_B$ regime (either due to extreme high $T$ or $\mu$), perturbative QCD (pQCD) becomes applicable, where asymptotic freedom ensures that quarks and gluons interact weakly~\cite{Ghiglieri:2020dpq,Vuorinen:2003fs,Andersen:2010wu,Haque:2014rua}. At $T$=0, the pQCD EoS has been computed at extremely large $n_B$ for zero isospin~\cite{Andersen:2002jz,Gorda:2021kme} as well as for largely asymmetric isospin \cite{Kurkela:2009gj}. Unfortunately, this regime is not directly relevant for astrophysics, however, pQCD calculations can provide constraints to models \cite{Komoltsev:2021jzg}. Combined QCD inferences can also provide information for the EoS in the cold dense regime~\cite{Fujimoto:2023unl}.

Experimental studies provide crucial constraints on the EoS at nuclear densities. Measurements of neutron skin thickness in heavy nuclei, obtained from, e.g., parity-violating electron scattering experiments, offer valuable insights into the symmetry energy~\cite{Suzuki:2022mow}. The  symmetry energy is a measure of the energy cost to make nuclear systems more neutron rich, or isospin asymmetric. Measurements of this quantity provide a way to use experimental data obtained on Earth to understand the interior of neutron stars.
Beyond $n_{\rm sat}$, new forms of matter become more energetically favored. Exotic particles such as hyperons and meson condensates can appear and alter the dense-matter EoS, generally softening it by opening more Fermi channels and decreasing the energy level in each, lowering the pressure for a given energy density and reducing the maximum neutron star mass, leading to the so-called hyperon puzzle~\cite{Bednarek:2011gd}. Possible resolutions to this problem include introducing hyperon-hyperon interactions~\cite{Rijken:2016uon}, three-body repulsive forces~\cite{Lonardoni:2014bwa}, or considering a phase transition to deconfined quark matter~\cite{Vidana:2005mx, Moss:2024uam, Fujimoto:2024doc}. At high energies, ultra-relativistic heavy-ion collisions also provide information about the EoS, including, e.g., evidence for quark deconfinement from collective flow~\cite{Ollitrault:1992bk}.

Furthermore, astrophysical observations have significantly refined the range of viable EoS models to describe dense matter. Accurate mass measurements of neutron stars exceeding two solar masses, radius estimates from the Neutron Star Interior Composition Explorer (NICER)~\cite{Miller:2021qha,Riley:2021pdl}, and tidal deformability constraints from gravitational wave events~\cite{LIGOScientific:2018hze,Chatziioannou:2020pqz} have imposed stringent limits on the stiffness of the EoS. These observations suggest that the EoS must be sufficiently stiff to support massive neutron stars, while still being soft enough to be consistent with radius and tidal deformability constraints.
At effectively $T$=$0$, neutron-star matter is governed by nuclear interactions, $\beta$-equilibrium, and charge neutrality. The Tolman-Oppenheimer-Volkoff (TOV) equations~\cite{Tolman:1939jz,Oppenheimer:1939ne} describe the structure of non-rotating neutron stars, yielding mass-radius relations that can be directly tested against observations. Neutron stars, whether in isolation or in binary systems, are governed by general relativity, requiring the EoS as an input to connect pressure to energy density.
The densities relevant for neutron star cores can reach several times $n_{\rm sat}$, making connections between theoretical calculations, terrestrial experiments, and astrophysical observations crucial. However, some extrapolation is necessary, and the conclusions drawn depend on the particular scheme chosen. Understanding the EoS thus requires a multi-faceted approach, combining lattice QCD, perturbative QCD, $\chi$EFT, nuclear experiments, and astrophysical observations.

In this proceeding, we briefly summarize the constraints on dense matter derived from first-principles calculations, experimental measurements, and astrophysical observations (see~\cite{MUSES:2023hyz} for the full review). We present two QCD phase diagrams, where we summarize our constraints, first independently of isospin, then in a new diagram as a function of isospin fraction.

\section{Constraints on  the QCD phase diagram as a function of net baryon density}

\begin{figure}
    \centering
    \includegraphics[scale=0.45]{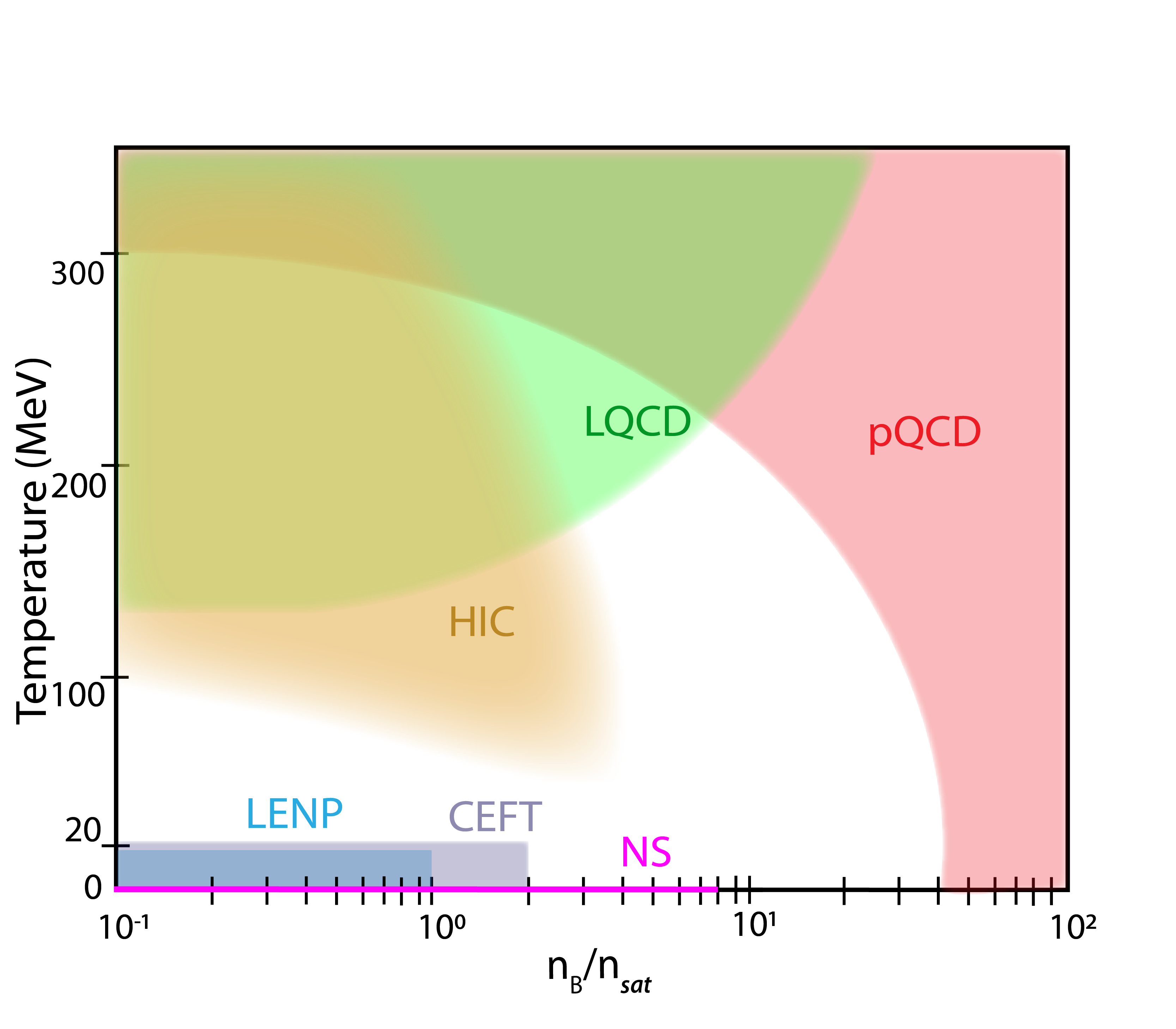}
    \caption{Constraints on different regions of the QCD phase diagram ($T$ vs. $n_B$) from various theoretical and experimental approaches: heavy-ion collisions (HIC), lattice QCD (LQCD), perturbative QCD (pQCD), low-energy nuclear physics (LENP), chiral effective field theory (CEFT), and astrophysical observations of neutron stars (NS). Figure taken from Ref.~\cite{MUSES:2023hyz}.}
    \label{fig:phase_diagram_nB}
\end{figure}

The phases of QCD (matter containing nuclei, bulk hadronic matter, and deconfined quark matter) are usually depicted in a diagram showing $\mu_B$ or $n_B$ vs. of $T$, the QCD phase diagram \cite{LRP1983}. The EoS of dense matter is constrained by several theoretical approaches, including  LQCD, pQCD, and $\chi$EFT, as well as from laboratory experiments and astrophysical observations. We now discuss each of these constraints and delimit their regime of validity across $T$ and $n_B$ in a (isospin independent) QCD phase diagram, \Cref{fig:phase_diagram_nB}.
In green, LQCD provides reliable results at high temperatures ($T \gtrsim 125$ MeV) and small baryon chemical potentials ($\mu_B \lesssim 300$ MeV). In pink, pQCD becomes valid at extremely high densities ($n_B \gtrsim 40~ n_{\rm sat}$) and/or high temperatures ($T \gtrsim 300$ MeV), where asymptotic freedom weakens the strong interaction. In gray, $\chi EFT$ or CEFT describes low-density nuclear matter at $n_B \lesssim 2~n_{\rm sat}$ and temperatures below $T \lesssim 20$ MeV, providing systematic uncertainty quantification for nuclear interactions. Heavy-ion collisions (HICs) in brown explore a broad range of temperatures ($T \sim 50-650$ MeV) and $n_B$ depending on the beam energy $\sqrt{s_{NN}}$, with high energies probing the quark-gluon plasma and lower energies accessing more baryon-rich hadronic matter. In blue, many different constraints extracted from low-energy nuclear physics (LENP) experiments near $n_{\mathrm{sat}}$ are combined. In pink, neutron star observations provide constraints for a wide range of $n_B$, in the low-temperature regime (effectively at $T$=0).

\subsection{Theoretical Constraints}

At vanishing or small baryon chemical potential ($\mu_B$=$0$), lattice QCD provides a reliable determination of the equation of state (EoS) for temperatures above approximately 125 MeV~\cite{Borsanyi:2013bia,HotQCD:2014kol}. These numerical calculations  reveal that the transition from a hadron resonance gas (HRG) to a quark-gluon plasma is a smooth crossover. 
LQCD has proven to be highly effective for investigating strong interactions in the vicinity of and beyond the deconfinement phase transition zone within the QCD phase diagram in the high-temperature and low-$\mu_B$ regime, primarily due to its ability to handle non-perturbative aspects~\cite{Ratti:2018ksb}. Extending to finite $\mu_B$, although no first-order phase transition has been found, a (chiral) pseudo-phase transition line can be calculated, based on where the phase transition order parameters change rapidly.
According to the latest LQCD results, the presence of a critical point (where a first-order phase transition coexistence line begins) is excluded for $\mu_B < 450$ MeV at a $2\sigma$ confidence level \cite{Borsanyi:2025dyp},
and the critical temperature (if a critical point exists) is expected to be smaller than $T^{\rm HQ}_c=132^{+3}_{-6}$ MeV for isospin-symmetric matter with zero strange ($\mu_S$) chemical potential~\cite{HotQCD:2019xnw}. 
Down that path, recent works employing different methods based on LQCD data have even been able to find evidence of the existence of a critical point within $400 \leq \mu_B \leq 650$ MeV \cite{Clarke:2024ugt, Shah:2024img}.

Based upon LQCD calculations, the chiral symmetry restoration crossover or pseudo-critical temperature at $\mu_B=0$ has been identified with high accuracy as $T^p_c=158\pm0.6$ MeV~\cite{Borsanyi:2020fev}. Additionally, a first-order deconfinement phase transition has been observed at $\mu_B=0$ for pure glue (QCD without quarks) at a temperature of $T_c^d\simeq 270$ MeV~\cite{Roessner:2006xn}. 
However, at finite $\mu_B$, the sign problem prevents direct Monte Carlo simulations, limiting the applicability of lattice QCD to chemical potentials up to $\mu_B\sim 3.5 T$~\cite{Borsanyi:2021sxv,Borsanyi:2022qlh}. 
Despite these challenges, lattice QCD can constrain the hadronic spectrum through partial pressures (with the hadronic
phase treated as ideal resonance gas)~\cite{Alba:2017mqu} and provide insights into strangeness-baryon interactions via correlations among the different conserved charges (baryon number $B$, strangeness $S$, and electric charge $Q$) given by derivatives of the pressure with respect to the different chemical potentials ($\mu_B$, $\mu_S$, and $\mu_Q$)~\cite{Borsanyi:2018grb,Bellwied:2019pxh}. Nevertheless, its reach does not extend to the high $\mu_B$ relevant for neutron stars.

At high temperatures and/or large chemical potentials, pQCD becomes applicable due to asymptotic freedom, where the strong coupling constant $\alpha_s$ decreases, and Debye screening reduces interactions among quarks and gluons~\cite{Gross:1973id}. Analytic calculations of the pQCD EoS are reliable above $T \gtrsim 300$ MeV at $\mu_B=0$ and at densities exceeding $n_B \gtrsim 40 n_{\rm sat}$ at $T=0$ \cite{Ghiglieri:2020dpq}. Theoretical improvements, such as resummation techniques using effective field theory \cite{Braaten:1995jr} and hard-thermal-loop perturbation theory \cite{Andersen:1999fw,Andersen:2002ey,Andersen:2003zk,Andersen:2009tc,Andersen:2011sf,Haque:2014rua}, have extended these calculations to next-to-next-to leading order (N2LO) at $\mu_B=0$ \cite{Andersen:2011sf} and partially to next-to-next-to-next-to leading order (N3LO) at finite chemical potential \cite{Freedman:1976xs}. Quantities provided include the EoS~\cite{Haque:2014rua}, the curvature of the QCD phase transition line~\cite{Haque:2020eyj}, and transport coefficients such as the sheer viscosity~\cite{Ghiglieri:2018dib}.
At T=0, the most up-to-date pQCD results for the EoS at finite densities can be found in \cite{Gorda:2021kme}. Although their applicability at $n_B \gtrsim 40 n_{\rm sat}$ is very far from the regime reached inside neutron stars, pQCD can still impose constraints on the EoS relevant for stellar interiors~\cite{Komoltsev:2021jzg}.

In the low-density regime ($T\lesssim20$ MeV, $n_B\lesssim 2 n_{\rm sat}$), $\chi$EFT provides a systematic, model-independent approach to describing hadronic interactions. Based on the symmetries of low-energy QCD, $\chi$EFT incorporates nucleons and pions as degrees of freedom and employs an order-by-order expansion constrained by empirical two- and few-body scattering data. One of its advantages over phenomenological models is the ability to quantify uncertainties through convergence studies and Bayesian methods \cite{Drischler:2021kxf}. $\chi$EFT provides the EoS computed up to N3LO in many-body
perturbation theory (with three-body
forces up to N2LO) \cite{Holt:2016pjb}. It also provides the symmetry energy and its slope (as a function of density) at $n_{\rm{sat}}$ \cite{Holt:2016pjb} and can be used to study the finite temperature behavior of nuclear matter by estimating the critical point for the nuclear liquid-gas phase transition for isospin-symmetric nuclear
matter (from a finite-temperature
calculation up to $T\sim25$ MeV \cite{Wellenhofer:2014hya}).

\subsection{Experimental Constraints}

Heavy-ion collisions provide a unique experimental avenue to probe the EoS of dense QCD matter at varying $T$ and $\mu_B$. Depending on the center-of-mass beam energy ($\sqrt{s_{NN}}$), these collisions explore a wide range of thermodynamic conditions, with high $\sqrt{s_{NN}}$ leading to high temperatures and lower baryon chemical potential, and low $\sqrt{s_{NN}}$ accessing higher $n_B$ at more moderate temperatures. The evolution of the system in these collisions follows a dynamic trajectory through the QCD phase diagram, where different experimental observables provide insights into distinct stages of this evolution.
For high $\sqrt{s_{NN}}$, the formation of a quark-gluon plasma (QGP) is expected, and hydrodynamic models describe the system well. In contrast, at very low $\sqrt{s_{NN}}$ (below $\sim 4-7$ GeV), the hadronic phase dominates, making hadronic transport models more suitable. The transition between these two regimes remains a topic of debate, and identifying the exact switching point is an ongoing challenge. The freeze-out temperature, which marks the point at which inelastic collisions cease and particle abundances are fixed, can be extracted from experimental data by assuming thermal equilibrium at freeze-out 
\cite{Hagedorn:1965st, Becattini:2001fg, Wheaton:2011rw, Petran:2013dva, Andronic:2018qqt, Vovchenko:2019pjl}. 
In contrast, the extraction of $n_B$ is more model-dependent. Emission of electromagnetic probes, such as photons and dileptons, which do not interact strongly, allows for the reconstruction of the temperature evolution of the system \cite{Strickland:1994rf,Gale:2025ome}.

An exciting frontier in heavy-ion collision research is the search for a possible QCD critical point. If such a critical point exists, susceptibilities of the pressure should diverge at that location in the phase diagram, leading to enhanced fluctuations of conserved charges, such as the kurtosis of the proton distribution. Measurements of these fluctuations have been carried out in the Beam Energy Scan (BES-I) program \cite{STAR:2020tga, STAR:2022qmt, HADES:2020wpc, ALICE:2019nbs}, with some early hints of non-monotonic behavior, though with large statistical uncertainties. 
Preliminary data from BES-II, coming with significantly improved precision, seem to confirm this non-monotonic behavior hinting for the potential existence of a critical point \cite{Stephanov:2024xkn}, although finalized results have not yet been released.
Further constraints on the EoS at high $\mu_B$ and moderate temperatures will come from the upcoming Compressed Baryonic Matter (CBM) experiment at FAIR, which will provide high-statistics measurements in fixed-target mode \cite{PANDA:2009yku}.

One of the most promising experimental observables for constraining the EoS at intermediate and low $\sqrt{s_{NN}}$ is the anisotropic flow of produced particles. These azimuthal anisotropies, known as flow harmonics, are sensitive to the pressure gradients in the system and, consequently, to the underlying EoS \cite{Danielewicz:2002pu}. However, there is considerable uncertainty in the correct dynamical model and transport coefficients used to extract the EoS from experimental data. Different modeling assumptions can lead to radically different posterior distributions for the EoS, as highlighted by studies comparing various approaches \cite{Danielewicz:2002pu}. As a result, while heavy-ion experiments provide crucial constraints on the EoS, extracting precise implications from the data remains an open challenge that requires further theoretical and computational advancements. 
Nevertheless, a step towards quantitative measurement of the speed of sound in HIC was recently achieved by the ALICE and CMS collaboration, although limited to single-temperature-valued results~\cite{CMS:2024sgx, ALICE:2024oox}.  

In the future, improved experimental precision, combined with more sophisticated dynamical models, will allow for a more direct extraction of the EoS from heavy-ion data. The ongoing and upcoming programs at RHIC, FAIR, NICA, and J-PARC will be instrumental in refining our understanding of dense QCD matter by systematically probing different regions of the QCD phase diagram.
On the other hand, low-energy nuclear physics experiments provide essential constraints on the by probing nuclear interactions near and at $n_{\mathrm{sat}}$. These experiments primarily investigate the properties of nuclei in regimes that overlap with neutron star crusts and the outer core, helping to bridge the gap between laboratory data and astrophysical observations.

At significantly lower beam energies, dense matter properties are studied through experiments that explore nuclei near $n_{\mathrm{sat}}$. While most stable nuclei consist of nearly symmetric nuclear matter with a charge fraction (number of protons over the number of nucleons) of $Y_Q=Z/A \sim 0.5$, heavier and neutron-rich nuclei can reach $Y_Q \sim 0.4$, and unstable nuclei near the neutron drip line have even smaller values. 
Key properties of symmetric nuclear matter, such as the saturation density, binding energy per nucleon, and compressibility serve as fundamental constraints on nuclear EoS models. The saturation density has been estimated to be $n_{sat}=0.17\pm0.03 ~\mathrm{fm}^{-3}$ from measurements of volume of a nucleus~\cite{Haensel:1981p}, while larger values of $n_{sat}=0.148-0.185~\mathrm{fm}^{-3}$ can be obtained within relativistic approaches \cite{Gross-Boelting:1998qhi,Blaizot:1980tw}. On the other hand, lower values $n_{sat}=0.1480\pm0.0038~\mathrm{fm}^{-3}$ came out of the PREX collaboration \cite{PREX:2021umo}. The binding energy per nucleon has been determined through mass measurements of heavy nuclei, with values of $B/A = -15.677$ MeV at $n_{\mathrm{sat}} = 0.16146$ fm$^{-3}$ \cite{Myers:1966zz} and $B/A = -16.24$ MeV at $n_{\mathrm{sat}} = 0.16114$ fm$^{-3}$ \cite{Myers:1995wx}. The nuclear incompressibility parameter ($K$), which quantifies the stiffness of nuclear matter, has been estimated from the Isovector Giant Monopole Resonance (ISGMR) in $^{90}$Zr and $^{208}$Pb, yielding a value of $K = 240 \pm 20$ MeV \cite{Colo:2013yta,Todd-Rutel:2005yzo,Colo:2004mj,Agrawal:2003xb}. However, a comprehensive review covering various experimental and theoretical approaches from 1961 to 2016 suggests a broader range of $K$ values between 100 MeV and 380 MeV~\cite{Stone:2014wza}. A more constrained range of $250 < K < 315$ MeV was determined by considering the Coulomb effect independently of microscopic models \cite{Stone:2014wza}.

The symmetry energy $E_{\rm sym}$ and its density-dependent slope parameter ($L$) are crucial for describing asymmetric nuclear matter and neutron stars. A comprehensive analysis of nuclear experiments and astrophysical data yielded fiducial values of $E_{\rm sym} = (31.6 \pm 2.7)$ MeV and $L = 58.9 \pm 16$ MeV \cite{Li:2019xxz}. The slope parameter, extracted from nuclear mass measurements, was found to be $L = 50.0 \pm 15.5$ MeV at $n_{\mathrm{sat}} = 0.16$ fm$^{-3}$ \cite{Fan:2014rha}. Meanwhile, the PREX-II experiment, which measured the neutron skin thickness of $^{208}$Pb, provided two different constraints on $L$: one study reported a high value of $L = 106 \pm 37$ MeV \cite{Reed:2021nqk}, while another, based on parity-violating asymmetry measurements, found a lower value of $L = 54 \pm 8$ MeV \cite{Reinhard:2021utv}, consistent with previous astrophysical estimates. 

Beyond nucleons, the properties of hyperons and $\Delta$-baryons in symmetric nuclear matter provide additional constraints on the EoS. The optical potential at nuclear $n_{\mathrm{sat}}$ is a key observable for effective model calibration. Experimental data from the 1980s determined the $\Lambda$-nucleon potential to be approximately $U_\Lambda \sim -28$ MeV \cite{Millener:1988hp}, with more recent estimates ranging from $-32$ to $-30$ MeV \cite{Fortin:2017dsj}. Measurements from the KEK facility in Japan suggest a repulsive potential for $\Sigma$ hyperons, with $U_\Sigma = 30 \pm 20$ MeV, while data from the KEK and J-PARC collaborations indicate an attractive potential for $\Xi$ hyperons, with $U_\Xi = -21.9 \pm 0.7$ MeV \cite{Gal:2016boi}. The ALICE collaboration’s studies of proton-$\Xi$ correlations suggest a less attractive $\Xi$ potential of $U_\Xi = -4$ MeV, consistent with lattice QCD calculations from the HAL-QCD collaboration, which report $U_\Xi = -4$ MeV, $U_\Lambda = -28$ MeV, and $U_\Sigma = +15$ MeV, with an uncertainty of approximately $\pm 2$ MeV \cite{Fabbietti:2020bfg,ALICE:2020mfd,Inoue:2019jme}.
It is worth mentioning that the recent measurements of proton-$\Omega$ correlations in proton-proton collisions by the ALICE collaboration  indicate an even more and purely attractive potential, but the analysis still needs to be pursed further for any possible quantitative comparison with LQCD prediction \cite{ALICE:2020mfd}.

Finally, at higher temperatures, nuclear experiments also probe the liquid-gas phase transition in hadronic matter. Data from heavy-ion collisions have provided information on the critical point of this transition \mbox{$T^{\rm LG}_c\simeq 15-17$ MeV}, \mbox{$ \mu^{\rm LG}_{B,c}\approx910$ MeV}, where nuclear clusters dissolve into bulk nuclear matter \cite{Elliott:2012nr}. Understanding these phase transitions is particularly relevant for studying the EoS at conditions encountered in, e.g., core-collapse supernovae and neutron star mergers.

\subsection{Observational Constraints}

Observational constraints on the EoS of dense matter are obtained from a combination of electromagnetic and gravitational-wave measurements of neutron stars. The macroscopic properties of neutron stars, such as maximum mass ($M_{\rm max}$), radius ($R_{M_{\rm max}}$), and tidal deformability ($\tilde{\Lambda}$), are intrinsically linked to the microphysics of nuclear matter. These properties provide critical benchmarks against which theoretical models of dense matter are tested.
Most importantly, the masses of neutron stars have accurately been constrained to be at least $\sim 2$ solar mass ($M_\odot$), PSR J0740+6620 with $M\geq 2.08 \pm 0.07$ $M_\odot$~\cite{Fonseca:2021wxt} and
PSR J0348+0432 with $M\geq 2.01 \pm 0.04$ $M_\odot$~\cite{Antoniadis:2013pzd}. Furthermore,
X-ray observations from NASA’s NICER have significantly improved radius measurements of neutron stars. For instance, the radius of a $2.072_{-0.066}^{+0.067}$ $M_\odot$ neutron star was determined to be $12.39_{-1.98}^{+1.30}$ km \cite{Miller:2019cac}, while a separate study found a radius of $13.7_{-1.5}^{+2.6}$ km for the same star, together with a mass of $2.08_{-0.07}^{+0.07} M_\odot$ \cite{Riley:2019yda}. These measurements constrain the mass-radius relation predicted by various EoS models.

Gravitational-wave observations from binary neutron star (BNS) mergers provide complementary constraints on the EoS, particularly through the tidal deformability parameter ($\tilde{\Lambda}$), which quantifies how much a neutron star deforms under its companion’s gravitational field. The event GW170817, detected by the Laser Interferometer Gravitational-wave Observatory (LIGO) and Virgo, placed an upper limit on the tidal deformability of neutron stars, constraining $\tilde{\Lambda} < 800$ for a 1.4 $M_\odot$ neutron star \cite{LIGOScientific:2018cki}. Such constraints are crucial in ruling out overly stiff or soft EoS models.

Beyond individual observations, multi-messenger astrophysics—combining data from electromagnetic emissions (radio and X-ray) and gravitational waves—provides a more complete picture of neutron star interiors. Pulsar timing from radio telescopes such as the Green Bank Telescope (GBT) \cite{Demorest:2010bx} yields the most precise neutron star mass measurements, setting a lower bound on $M_{\rm max}$, which any viable EoS must support. Meanwhile, the aftermath of neutron star mergers, including kilonova light curves and post-merger gravitational waves, informs the behavior of matter at extreme densities and temperatures \cite{Rezzolla:2017aly}. Future gravitational-wave detectors are expected to probe post-merger signals more effectively, which could provide further insights into the possible deconfinement of quark matter in neutron star cores \cite{Blacker:2020nlq,Magnall:2024ffd}.

\section{Constraints on  the QCD phase diagram as a function of isospin fraction}

\begin{figure}[t!]
    \centering
    \includegraphics[scale=0.7]{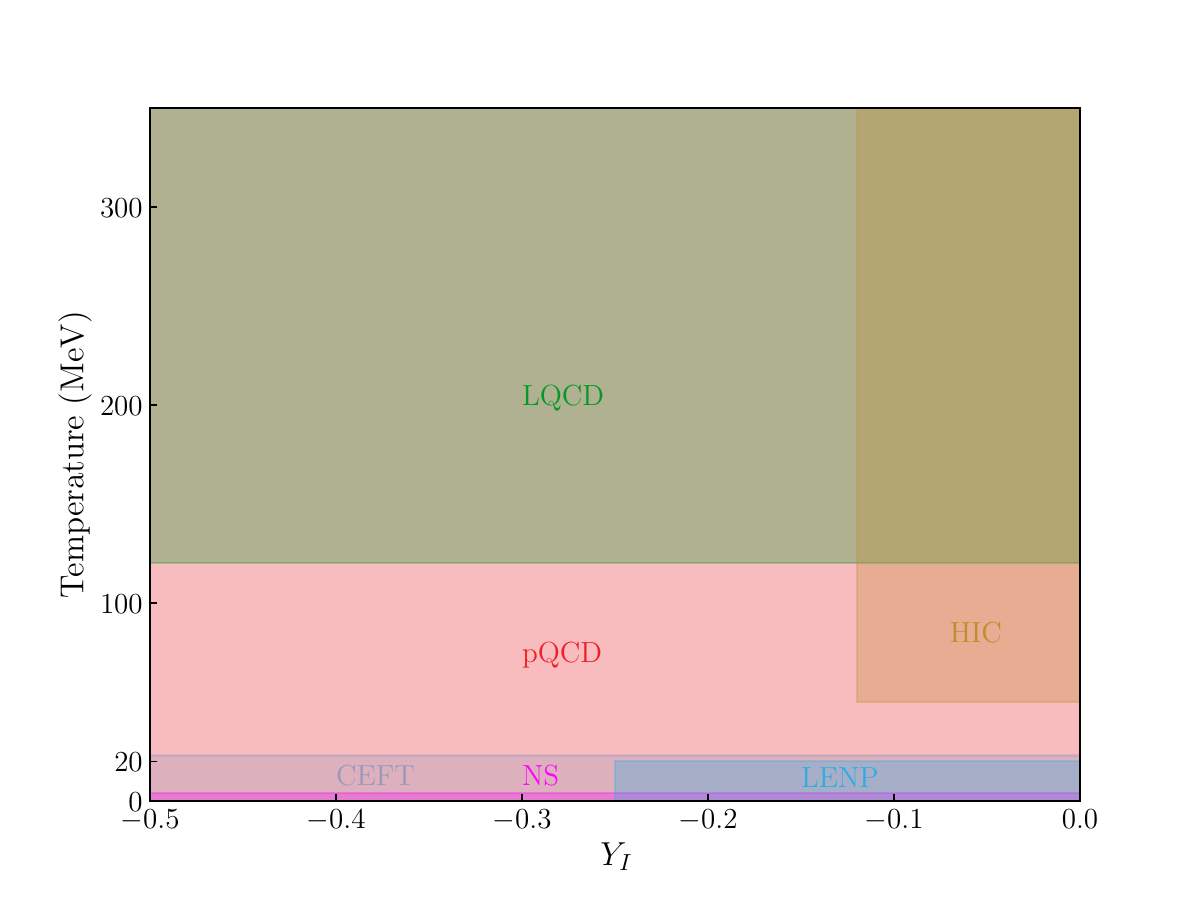}
    \caption{Same as \Cref{fig:phase_diagram_nB} but as a function of isospin fraction $Y_I$. It is implied that these regions do not necessarily correspond to the entire baryon density region covered in \Cref{fig:phase_diagram_nB}. Note that the pQCD region extends behind the other regions and covers the whole phase diagram.}
    \label{fig:phase_diagram_YI}
\end{figure}

In addition to exploring the QCD phase diagram as a function of $n_B$ and $T$, it is also essential to examine it as a function of some quantity related to isospin, Here we choose the isospin fraction $Y_I = {\sum_i I_{3,i} n_i}/{ n_B}$, where $I_3$ is the third component of isospin and $n$ the density of each particle. This quantity characterizes the relative abundance of different particles in isospin multiplets in dense matter and is crucial in describing neutron-rich astrophysical systems such as neutron stars and their mergers.
It relates to the aforementioned charge fraction as $Y_I=Y_Q-\frac1 2 + \frac 1 2 Y_S$~\cite{Aryal:2020ocm}, where the charge fraction can also be defined as $Y_Q = {\sum_i Q_{i} n_i}/{ n_B}$, and the strangeness fraction as $Y_S = {\sum_i S_{i} n_i}/{ n_B}$. \Cref{fig:phase_diagram_YI} illustrates the same theoretical and experimental constraints already discussed and plotted in Figure 1, but now shown as a function of $Y_I$. The figure highlights different regions of interest, including symmetric nuclear matter ($Y_I=0$) on the right, relevant for nuclei, and pure neutron matter on the left ($Y_I =-0.5$), which is more relevant for astrophysics. The numbers associated with the different colored regions are described in Table~1.

At high $T$ and moderate $n_B$, LQCD calculations have been used to study the thermodynamic properties of QCD matter, providing insight into the deconfinement transition and critical behavior at different charge asymmetries. The expansion already shown in Fig.~1 has not yet been extended to isospin asymmetric matter, but an expansion until lower $\mu_B/T=2.5$ is valid for any isospin fraction~\cite{Guenther:2017hnx}. In particular, Ref.~\cite{Guenther:2017hnx} tuned $\mu_S(\mu_B, T)$ and $\mu_Q(\mu_B,T)$ to enforce strangeness neutrality and electric charge conservation to reproduce the conditions $Y_S=0$ and $Y_Q=0.4$. This is possible because LQCD does not suffer from the sign problem concerning isospin unbalance (as it does for baryon/antibaryon unbalance)~\cite{Brandt:2017oyy}.

In the high-energy regime, pQCD calculations provide constraints on the EoS at extreme $T$ and/or $\mu_B$, where hadronic matter transitions into deconfined quark matter, then into weakly interacting quark matter. Extensions to arbitrary isospin can be found for any $T$ and $\mu_B$ combinations,
\begin{wraptable}{r}{0.34\textwidth}
    \label{tab:yi_regions}
\begin{tabular}{|c|c|}
        \hline
        Region & Isospin Fraction ($Y_I$) \\
        \hline
        \hline
        PQCD    & $-0.5$ to $0$ \\
        \hline
        LQCD    & $-0.5$ to $0$ \\
        \hline
        CEFT   & $-0.5$ to $0$ \\
        \hline
        NS     & $-0.5$ to $0$\\
        \hline
        LENP    & $(-0.5)$, $-0.25$ to $0$ \\
        \hline
        HIC    & $-0.12$ to $0$ \\
        \hline
    \end{tabular}
        \caption{Region of applicability of the theories and experiments shown in the figures with respect of isospin.}
\vspace{-.3cm}
\end{wraptable} 
as long as $T$ and/or $\mu_i$ for the different quarks correspond to high energy \cite{Vuorinen:2003fs,Haque:2014rua}.
See Ref.~\cite{Brown:2024gqu} for an idea of how different chemical potentials and constraints affect how matter reach the conformal (non-interacting massless Fermi gas of quarks) limit.

At low $T$, for $n_B\sim n_{\rm{sat}}$, isospin-dependent effects are well-described by $\chi$EFT, which can be used to study the EoS at arbitrary isospin asymmetry~\cite{Drischler:2013iza,Wellenhofer:2016lnl,Wen:2020nqs,Somasundaram:2020chb}. However, in practice it is convenient to first compute the EoS for symmetric nuclear matter ($Y_Q=0.5$) and pure neutron matter ($Y_Q=0$) and then interpolate between the two using the symmetry energy expansion \cite{Bombaci:1991zz}.

Heavy-ion collisions collide nuclei that are either isospin asymmetric ($Y_I=-0$) or slightly asymmetric (e.g., Uranium $Y_I=-0.12$), although  isospin fluctuations remain finite during the evolution \cite{Plumberg:2024leb, Nana:2024okk}.
With the exception of experiments that focus on neutron matter ($Y_I=0$) (e.g., tetraneutron experiments~\cite{Duer:2022ehf}), low-energy nuclear experiments produce matter that is either isospin symmetric ($Y_I=0$), or asymmetric up to $Y_I=-0.25$ (e.g., He-8) at FRIB.
Neutron stars, on the other hand possess a variety of isospins, from isospin symmetric nuclei (such as Carbon) in the bottom of their inner  crusts, to very asymmetric matter in their cores ($Y_I\sim-0.45$).

\section{Conclusion}

In this proceeding we review current constraints on dense matter, including constraints from  heavy-ion collisions (HIC), lattice QCD (LQCD), perturbative QCD (pQCD), low-energy nuclear physics (LENP), chiral effective field theory ($\chi$EFT), and astrophysical observations of neutron stars (NS). 
In particular, recent theoretical results seem to narrow down the window for the existence of a critical point, with several predictions from rather different approaches. Experimental results from intermediate to low energy also help to infer with more precision the region of potential existence of this critical point, with important results awaited from BES-II, considering exciting preliminary results presented last year.
The emergence of multi-messenger astronomy in the past decade also contributed to improve our knowledge at very high-density, through more accurate measurements of neutron star properties and their merger.
Beyond that QCD phase diagram already discussed in \cite{MUSES:2023hyz}, we present here a new one where the different constraints are shown also taking into account isospin. In this new diagram, some regions are narrow in isospin fraction, e.g., heavy-ion collisions, and some are wide, e.g., lattice QCD and perturbative QCD. Neutrons stars have isospin fractions varying from close to -0.5 to 0, going from their centers to their surfaces.
This compilation allows us to form a complete map of the QCD phase diagram, for the first time quantitatively as a function of temperature, baryon density, and isospin fraction, connecting our current knowledge from theory and experiment, helping us understanding better dense QCD matter.

\bibliographystyle{apsrev4-1}

{\small
\bibliography{inspire,NOTinspire}
}

\end{document}